# The Second Law as a Cause of the Evolution

Oded Kafri

Varicom Communications, Tel Aviv 68165 Israel

**Abstract**

It is a common belief that in any environment where life is possible, life will be generated. Here it is suggested that the cause for a spontaneous generation of complex systems is probability driven processes. Based on equilibrium thermodynamics, it is argued that in low occupation number statistical systems, the second law of thermodynamics yields an increase of thermal entropy and a canonic energy distribution. However, in high occupation number statistical systems, the same law for the same reasons yields an increase of information and a Benford's law/power-law energy distribution. It is therefore, plausible, that eventually the heat death is not necessarily the end of the universe.

**Introduction**

Till the late 17th century the common hypothesis about the origin of life was Abiogenesis or a spontaneous generation of life from organic substances. For example, if we put an orange on a table, after a while we will probably find worms in it. The conclusion that the worms are generated from the orange is only part of the truth. We know today that eggs of a fly called *Drosophila melanogaster* have to be present in the orange, in the first place in order for the worms to be generated. Nevertheless, the "explanation" that the eggs are the reason for the generation of the worms in the orange does not diminish the appeal of the Abiogenesis hypothesis. For an observer in space looking at the earth over billions of years, the formation of life, buildings, roads, cars etc, cannot be explained in a similar way as the formation of worms in the orange. For him, the explanation for life and even buildings, books etc. is a spontaneous generation. What are the "eggs" of these objects?

Contemporary science deals with evolution, namely how life evolved from the simple to the complex. Nevertheless, evolution theories do not deal and do not explain the reason for spontaneous generation of complex objects as life and artificial objects [1].

Sometimes life is regarded as a thriving for order. It seems we are constantly fighting against the chaos invading our life, constantly looking for rules and laws, and if we don't find them we invent them. However, there is no scientific definition of order. It seems odd that order is not defined in science, while entropy, the entity that is conceived as a measure of disorder, is so widely used. According to the 2nd law of thermodynamics, any process that causes the entropy to increase is likely to happen. In other words, such processes are spontaneous. This is the reason for the bad reputation of the second law "which fights our tendency to order". The Internet is loaded with theological material claiming that life is a violation of the second law of thermodynamics.

In this paper it is argued exactly the opposite: The second law is responsible for life because life is related to a spontaneous information increase, and information is a part of entropy. It is argued that entropy comprises of two parts: the thermal entropy and the informatics entropy. While the thermal entropy and its connection to the $2^{nd}$ law are well known, the connection between information and the $2^{nd}$ law was only recently discussed [2,3,4].

Information is conceived (erroneously) as order. It was suggested that information contains null or even negative entropy [5]. This "intuitive reasoning" is logically sound, as many understand information as the requirement that Alice will obtain the same value for the $i^{th}$ bit each time Bob sends her an identical file. Since the bits location is fixed, the file is a frozen entity and thus contains zero entropy. Even more popular intuition is that information is negative entropy as was suggested by Brillouin [6,7]. The reasoning of this conclusion is that the file, before it is received, is unknown, and thus contains entropy, which is reduced with every bit that Alice reads. Therefore, information is a reduction of entropy.

Nevertheless, Shannon, in his $1^{st}$ theorem, defined the information as the maximum amount of data that can be transmitted in a noiseless channel. His expression, which is identical to the Gibbs entropy, represents the randomness of the distribution of the bits in a file [8]. The Shannon information tells us nothing about the actual content of a file and has no connection to it. When Alice receives a file of $\Lambda$ bits, all we know is that there are maximum $2^{\Lambda}$ different possible (configurations) contents. In some of these $2^{\Lambda}$ files the distribution of the bits is correlated. In others the distribution of the bits is random. When Alice receives a $\Lambda$ bits random file, the amount of the Shannon information is $I = \Lambda$ bits. If the distribution of the bits is not random Bob can compress the file and send a shorter file of length $I$ such that $I < \Lambda$ or in general $I \leq \Lambda$.

How many random distributions are there as compared with correlated distributions in a file? Jaynes has shown [15,19,20] that if we have a statistical ensemble, the most honest guess about its distribution is the Shannon information. The finding of Jaynes, in simple words, is that there are much more random distributions than there are correlated distributions. The work of Jaynes suggests that for a statistical ensemble the $2^{nd}$ law is a mere probabilistic effect. If we don't know anything about a statistical ensemble, our best bet is that it is random. If we reshuffle a distribution of something, it will become more random.

Jaynes work is applicable to information. If we add a noise to a file, it will probably increase the amount of the Shannon information in the file. Nevertheless the subjective meaning of the message in the file may be lost.

While the similarity between the Gibbs entropy and the Shannon information is clear, there is a distinctive difference between them. Information is a logical quantity; the Shannon information is a mathematical entity. It is neither a function of the file energy nor a function of the temperature, as it is not made of a materialistic substance. A binary information file contains only "1"'s and "0"s. Entropy on the other hand, is a physical quantity and has a physical dimension. The physical meaning of the entropy is derived from the second law. The basic outcome of the second law is that heat flows spontaneously from a hot bath to a cold bath; it does so, because in this process the entropy increases.

Another difference is that entropy is a dynamic quantity while information is a quenched quantity. Boltzmann obtained his approximation of the Clausius entropy for an ideal gas, which contains a large number of atoms exchanging energy constantly [12]. The Boltzmann statistics contains inherently the canonic non-deterministic Maxwell-Boltzmann distribution. This distribution is a cornerstone of mechanical statistics as well as of quantum mechanics. Nevertheless, the canonic distribution is not applicable to information, which is a quenched quantity.

**I. The paper in a nutshell**

In previous publications [2,3] it was shown that if we assign energy to the information bits, it is obtained, from the $2^{nd}$ law of thermodynamics, that the Shannon information is entropy, a random file is a state of equilibrium and the temperature is proportional to the bit's energy. In this paper a toy model is used to describe a generic file transmission from Bob to Alice using electromagnetic radiation. Bob is using a blackbody-based transmitter that enables him to control the temperature and the frequency of the radiation. In addition, Bob can modulate the radiation. It is shown that when Bob transmits to Alice a low occupation number energy (the quantum limit), the thermodynamic functions of the energy transmission are the well-known canonic ones.

However, when Bob increases the occupation number of the photons, a power-law distribution replaces the canonic distribution, information replaces the entropy, and the canonic statistical physics becomes a statistics of harmonic oscillators. In the high occupation limit the obtained normalized thermodynamic functions are independent of any physical quantity and/or physical constant and therefore become purely logical functions. A qualitative discussion about whether nature prefers generation of information or generation of canonic entropy yields a conclusion that the two are equally welcome.

In section **II,** *a toy model,* in which Bob sends a collimated light beam to Alice, is described. Bob is using a blackbody radiation source that delivers energy per mode according to the Bose-Einstein statistics. Bob can change to his wish (without any physical limitations) the temperature of his source and to select a frequency or frequencies of the radiation by a spectral filter. In addition, Bob has a shutter that enables him to modulate the radiation within the limitations of the laws of optics. It is assumed that in equilibrium all the radiation modes that Alice receives have the same temperature.

In section **III,** *the quantum limit*, it is assumed that the energy of the photons is much higher than the average energy. The Bose-Einstein equation yields the familiar canonic distribution. The obtained entropy of the radiation is the Gibbs expression and the ratio between the number of the photons and the number of empty modes is the Maxwell-Boltzmann distribution.

In section **IV**, *the high occupation limit*, Bob is using his toy model to reduce the energy of the photon as compared with that of the average energy of the mode to the extent that the energy can be added or be removed smoothly to the mode. In this limit the number of the photons is much larger than the number of the modes. The Bose-Einstein equation yields that each mode is a harmonic oscillator. It means, that each mode's entropy is one Boltzmann constant, and the temperature is a linear function of the mode's energy.

In section **IV-a,** *modulation and information*, Bob is modulating the sequence of the harmonic oscillators to a binary file. The Shannon information is calculated. It is shown that in a random file, when the number of the harmonic oscillators (energetic modes) is equal to the number of the vacancies (empty modes), the Shannon information is equal to the length of the file. In other cases it is shown that the amount of the Shannon information is smaller.

In section **IV-b,** *equilibrium and entropy,* the entropy of the file, which consists of harmonics oscillators and vacancies, is calculated. It is shown that the Shannon information is the Gibbs mixing entropy. The Boltzmann $H$ function is equivalent to the amount of information $H$ in a correlated file.

In section **IV-c,** *logical quantities,* it is shown that the normalized entropy in Bob's transmission is a function that does not contain any physical variable or constant. This is with contradistinction to a canonic entropy transmission. Therefore the normalized entropy, in the high occupation limit, is a logical quantity.

In section **IV-d,** *logical equilibrium –the Benford's law,* Bob generates a set of modes in which each represents a digit. A possible way to construct such a set is to put in the mode that represents the digit $N$, $N$ times more energy than the mode that represents the digit 1. To obtain equilibrium (equal temperature in the Bose-Einstein distribution) Bob has to use either a different frequency for each digit-mode or alternatively a different density for each digit in a file. The obtained normalized distribution function of the digit-modes in equilibrium is a pure logical function, which is not a function of the initial temperature and/or frequency chosen by Bob. The result is identical to the famous Benford's law.

In section **IV-e,** *energy distribution, the power-law*, the log of the occupation number vs. the log of the photon energy divided by the average energy is plotted for the Bose-Einstein distribution. It is seen that in the low occupation number, a straight line of slop, $-1$, is obtained. This behavior appears in many quantities in nature (a good example are the natural nets). With analogy to the momentum Gaussian distribution obtained from the canonic energy distribution If we consider the electric field of the radiation instead of the energy, a $-2$ slope is obtained. Slopes around "-2" appear in many sociological statistics [20].

In section **V,** *mixed systems*, the question whether information can survive side by side with the canonic thermal entropy in equilibrium is discussed.

In section **V-a,** *Hook-law harmonic oscillator*, a thermodynamic analysis of a mixed system consists of a single Hook-law oscillator coupled to a heat bath at room temperature is presented. It is shown that the amplitude increase of a Hook-law oscillator has the Carnot efficiency, similar to that of the amplification of a file [3,4]. A Hook-law oscillator will relax its energy spontaneously to the heat bath as the relaxation increases the entropy.

In section **V-b,** *information vs. thermal entropy***,** it is shown that in the case of the blackbody emission, both the low occupation number photons as well as the high occupation number photons coexist in equilibrium. Therefore, it is concluded that information and thermal entropy are equally welcome.

In section **VI,** *summary and discussion,* a table that shows the differences between the thermodynamic functions in the canonic distribution and in the high-occupation harmonic distribution is presented. In view of these differences, the meaning of the logical quantities obtained in the thermodynamic theory of communication is discussed. It is concluded that, in equilibrium, inert quanta distributed in modes yield a power-law/Bedford-law distribution. It is suggested that the informatics aspect of life is a tendency for reproduction and a compressed communication.

**II. The Toy Model**

In Fig. 1 the setup in which Bob sends Alice a flux of photons is described. The analysis is based on the classical Carnot Clausius thermodynamic. In the classical thermodynamic the entropy is $S \geq Q/T$ where the equality sign stands for equilibrium, $Q$ is the heat that Bob is sending or Alice is receiving, and $T$ is the temperature of the transmitter or the receiver. This inequality is the Clausius inequality [13] derived directly from the efficiency of the Carnot cycle [12].

Bob is sending a sequence of photons in a single longitudinal mode to Alice, as described in Fig.1. Bob has a blackbody at temperature $T_H$ that emits a blackbody radiation. Bob attaches a pinhole filter (PH) of a diameter of $\lambda^2$ with a positive lens in order to obtain a collimated single longitudinal mode. After the pinhole spatial filter Bob attaches a spectral filer and a polarizer (SF) that passes only the frequency v, with a spectral width $\Delta v$. Here λ and v, are the wavelength and the frequency of the transmitted signal. The spectral width determines the number of the temporal modes. Bob can modulate the photons beam by using a mechanical or electro-optical shatter. Photons are bosons with a zero chemical potential. Therefore the number of photons n in a single mode obtained from the Bose-Einstein distribution [14] is given by;

$$n_i = \frac{1}{e^{hv/k_B T_H} - 1} \qquad (1)$$

Where i index the temporal modes, hv is the energy of the photon, $T_H$ is the temperature of the Bob's source and $k_B$ is the Boltzmann constant.

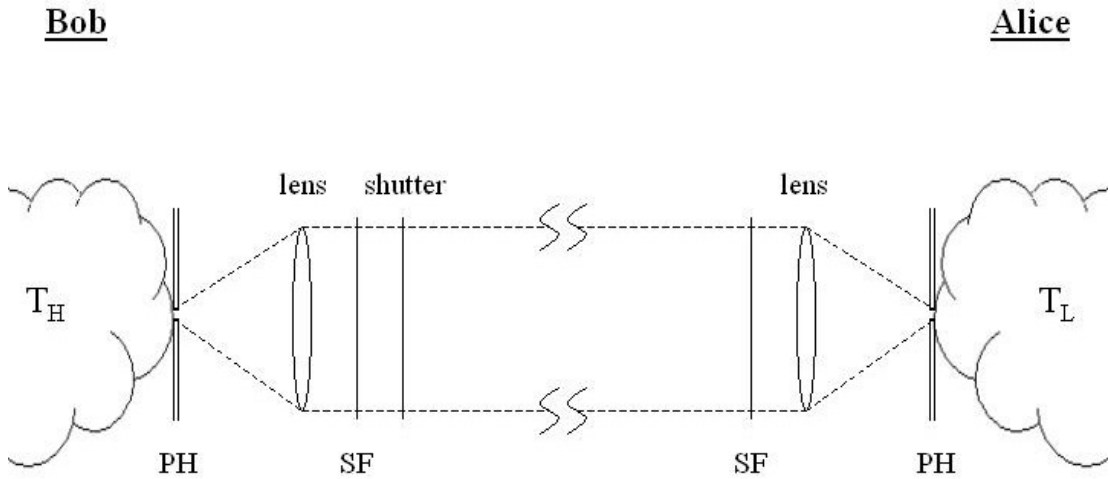

Fig 1. *A setup for a single transverse mode energy transmission from Bob to Alice*

Alice uses a detector to receive the message. In general, Alice uses a similar positive lens, a filter and a detector at the focal length of the lens. If the detector of Alice is at a temperature $T_L = T_H$, namely the temperature of the transmitter of Bob, the noise emitted by the detector will be as strong as the signal, and Alice will not be able to read the signals. Therefore, a prerequisite requirement for energy transmission from Bob to Alice is that $T_H > T_L$. In practice, Bob can heat his blackbody to a temperature limited by the physical properties of the blackbody's material. However, we assume that Bob does not have such limitations and he can produce a beam as hot as a laser beam to his wish. Also there are no limitations of the frequencies that Bob can send. In practice, the wavelength of the radiation cannot exceed the diameter of the blackbody; nevertheless we let Bob enjoy the benefit of a toy model. Hereafter, two limits are discussed: the quantum limit in which $h\nu \gg k_B T$ and the high occupation limit in which $h\nu \ll k_B T$.

**III. The Quantum Limit**

In the quantum limit, the energy of the photon $h\nu$ is much higher than the average energy, $k_B T$, of a mode emitted from a thermal bath (the blackbody). Therefore classically it is impossible to emit a photon. However, when many modes are collecting their energies together they emit a single high-energy photon in an arbitrary (lucky) mode. In other words, when $n \ll 1$, it is assumed that a group of $1/n$ modes will emit a single photon in an unknown mode of the group. Occupation numbers smaller than one exist in many systems in physics i.e. in ideal gas. In this case Eq(1) yields,

$$n_i = e^{-h\nu/k_B T_i}, \qquad (2)$$

which is the canonic distribution. Consider a sequence of $\Lambda$ temporal modes emitted from a radiation source, which is not in equilibrium. Non-equilibrium state means that any mode may have its own temperature. The number of photons in the sequence is $\sum_{i=1}^{\Lambda} n_i$. The average energy of a single mode is $q_i = n_i h\nu$. The temperature is calculated from Eq.(2) to be, $T_i = -h\nu/k_B \ln n_i$. The entropy of a single mode will be $S_i = q_i/T_i$ or, $-k_B n_i \ln n_i$. Since the entropy is extensive, the total entropy of a sequence of $\Lambda$ modes is,

$$S = \sum_{i=1}^{\Lambda} S_i = -k_B \sum_{i=1}^{\Lambda} n_i \ln n_i. \qquad (3)$$

When $n_i \ll 1$, $n_i = p_i$ and Eq.(3) is simply the Gibbs entropy. Assuming that all $n_i$ are equal to $n$ (which means an equilibrium state as all the temperatures $T_i$ are equal to $T$), we obtain from Eqs(3&2);

$$S = \sum_{i=1}^{\Lambda} S_i = \frac{\Lambda h\nu}{T} e^{-\frac{h\nu}{k_B T}} = \frac{\Lambda n h\nu}{T} = \frac{\Lambda q}{T}, \qquad (4)$$

The entropy of the sequence of $\Lambda$ temporal modes, in the quantum limit, is a function of the mode energy and its temperature. Any loss of a photon will change the entropy as well as any fluctuation of the source temperature.

**IV. The High Occupation Limit**

When the source is hot and/or the frequency is low such that $h\nu \ll k_B T$, Eq(1) yields,

$$n_i h\nu = q_i = k_B T_i \quad \text{or} \quad S_i = k_B \qquad (5)$$

This is the well-known relation of a harmonic oscillator. In this limit the photon energy is negligible as compared with the average energy of the mode, and therefore energy can be removed or added in a continuous way. The uncertainty of ½ $hv$ is also negligible. To some extent it is a surprising result. Intuitively, we expect from one mode, which contains many photons, to have zero entropy. Nevertheless, one $k_B$ is a very small amount of entropy, i.e. a laser mode, which sometimes contains as much as $10^{20}$ photons, has the same amount of entropy as one vibration mode of a single molecule. The Gibbs, Boltzmann or Von-Neumann entropies yield null entropy for a single mode radiation, as they are only an approximation of the entropy for a statistical ensemble [9]. Finite entropy to a single mode is a must, as entropy is an extensive quantity and the emission of entropy by a blackbody radiation is the sum of the single modes emission. Therefore if a single mode would not carry entropy, a blackbody would not emit entropy as well.

When Bob is using his Blackbody (in this case he will prefer a CW laser) and sending Alice $\Lambda$ classical modes, the total entropy that is removed from his source is,

$$S = \sum_{i=1}^{\Lambda} S_i = \Lambda k_B, \qquad (6)$$

In the next two sections it will be shown that entropy in the high occupation limit is not a simple sum of the entropies of the modes. A sequence of $\Lambda$ oscillators is not always a state of equilibrium since we can add as many empty modes as we wish. Therefore, Eq.(6) is a lower bound of the entropy. The entropy is defined only in equilibrium. The equation of the entropy can be used away from equilibrium, however the obtained value (usually known as the Boltzmann -*H* function) is not unique and is always smaller than the entropy [15].

The entropy of a single oscillator in the high occupation limit is not a function of the energy or of the temperature. In fact, when a mode is divided to *N* fractions, each fraction, when received, carries the same amount of entropy as the undivided mode. It was shown previously [2,3] that this property is the basis of information transmission and is the cornerstone of IT.

**IV-a. Modulation and Information**

A possible way for Bob to modulate his source is to use a shutter (Fig. 1). Every temporal mode has a duration of its coherence length, namely $\Delta t = c/\Delta v$. Where $c$ is the speed of light. Therefore if the shutter is opened for a time interval $\Delta t$, an amount $k_B$ of entropy is transmitted.

When Bob is transmitting a file he possibly starts by sending a header to inform Alice about the file length $\Lambda$ that he intends to send and some other information about the kind of compression or the language he uses. Usually Alice replies to confirm the acceptance of the header and her consent or refusal to receive the data and so on. However, although this handshaking process is vital to any communication, it will not be discussed here. The discussion here assumes that Bob and Alice have pre-agreed language, compression, protocol and an open channel of communication.

If Bob sends $L$ energetic bits in $\Lambda$ modes where, $\Lambda > L$, there are several different messages that can be sent. The number of possible messages is the binomial coefficient $\Lambda!/(\Lambda-L)! L!$. The Shannon information is defined, in bits, as the shortest file *I* that has this number of messages. Therefore, $2^I = \Lambda!/(\Lambda-L)! L!$. Hereafter the information will be expressed in nats, namely, $e^I = \Lambda!/(\Lambda-L)! L!$. Stirling formula yields,

$$I = \Lambda \ln \Lambda - L \ln L - (\Lambda-L)\ln(\Lambda-L). \tag{7}$$

Under the assumption that all the combinations have equal probability, it is seen that if $\Lambda = 2L$ then $I = \Lambda \ln 2$. Namely, a random file contains the maximum amount of information.

**IV-b. Equilibrium and Entropy**

The basic definition of equilibrium is obtained from Clausius inequality, namely $S \geq Q/T$. When the heat transmitted divided by the temperature is equal to the entropy, the system is in equilibrium. This implies that when $Q/T$ is a maximum, the system is in equilibrium. If the system is not in equilibrium, the obtained temperature (that is always higher than *T*) is not unique and can yield different values for different histories of a system.

If we designate $p \equiv L/\Lambda$, then the RHS of Eq.(7) can be rewritten as,

$$I = -\Lambda \{ p \ln p + (1-p)\ln(1-p) \} \tag{8}$$

Consider $p\Lambda$ oscillators, each carries $k_B$ entropy, in a sequence of $\Lambda$ modes. In the setup of Fig 1, each mode has the same frequency and temperature. That means that each mode is in equilibrium with the emitting Blackbody and with the other modes. However, there is mixing entropy of the energetic modes and the empty modes. The mixing entropy of the ensemble *a la* Gibbs is,

$$H = k_B \Lambda \{p \ln p + (1-p) \ln(1-p)\} \quad (9)$$

where $H$ is the Boltzmann $H$ function. The $-H$ function is the entropy calculated away from equilibrium such that $S \geq -H$. In equilibrium $p = \frac{1}{2}$, $H$ has a minimum and Eq.(8) yields that $S = k_B \Lambda \ln 2$.

Therefore it is possible to conclude that entropy and information are identical and a random sequence is a state of equilibrium. In the case that $p < \frac{1}{2}$ one obtains,

$$S \geq -H = k_B I \quad (10)$$

Eq.(10) is the Clausius inequality for informatics.

It worth noting that Eq.(9) yields, in statistical physics, the canonic distribution [2,3]. Consider $p\Lambda$ energetic particles of energy $h\nu$ in $\Lambda$ microstates. The energy of the sequence is $Q = p\Lambda\, h\nu$. The temperature is defined *a la* Clausius as $T = \partial Q/\partial S$, Therefore,

$\partial S/\partial p = -\Lambda h\nu/T = \Lambda k_B \{\ln p - \ln(1-p)\}$ or $p/(1-p) = e^{-h\nu/k_B T}$ which is the canonic distribution of Eq.(2) for a two level system (see table 1) [2,3].

What is the reason for such different results, in statistical physics and in informatics, obtained from the same Eq.(9)? The explanation is that in statistical physics, the canonic distribution prevails and the entropy of a mode is a function of the energy and the temperature, as is seen in Eq(4). Therefore, equilibrium (a maximum entropy) may be obtained, for any value of $h\nu$ and $T$, for any $p \leq \frac{1}{2}$. In informatics the entropy of a mode is not a function of the energy or the temperature, therefore equilibrium exists only for a single value $p = \frac{1}{2}$.

**IV-c. Logical Quantities**

When Bob modulates the transmitted radiation of the setup of Fig.(1), he usually does not care about the coherence length of his radiation source. He transmits a sequence of pulses and vacancies of equal length. Therefore, each "1" bit will usually carry more entropy than one $k_B$. Practically it will carry $K = m k_B$ entropy units, were $m$ is some integer (see Eq.(6)), therefore Eq.(10) can be rewritten as,

$$S \geq KI, \qquad (11)$$

When Bob transmits a file, all he wants is for Alice to receive correctly one of the $2^\Lambda$ possible files in a $\Lambda$ bits transmission. However, here **we** are interested in the amount of the Shannon information of this particular transmission. A possible way to calculate the amount of information in the transmission is to use Eq.(9) to calculate $-H/S$ which is the normalized information. Which yields,

$$-\frac{H}{S} = -\frac{p \ln p + (1-p) \ln(1-p)}{\ln 2} \leq 1. \qquad (12)$$

Eq.(12) is the logical Clausius inequality in informatics. It says that the maximum amount of information in a file that has a fraction $p$ of the bits "1" or "0" is not a function of $K$. In fact, Eq.(12) is a mere inequality, free from any physical quantity.

### IV-d. Logical Equilibrium – The Benford's Law

Eq.(12) demonstrates that the fraction $p$ of the "1" or "0" bits determines how far a file is from equilibrium. If $p=1/2$, it means that a file might be in equilibrium. Nevertheless, information transmission is not done usually in bits. In our everyday life, we are using a much larger amount of symbols to communicate. An important set of symbols is the numerical digits. A possible way to form a set that represents the nine digits is to use nine kinds of modes. Each one contains 1,2,3,4,5,6,7,8,9 energy units respectively. What will be the relative distribution of these modes in equilibrium? If all the bosons have the same energy, each occupation number $n$ yields a different temperature, which means a non-equilibrium state. Eq.(1) is rewritten as,

$$\Phi(n) = \Phi = \frac{h\nu}{k_B T} = \ln(1+\frac{1}{n}) \qquad (13)$$

It is seen that as $n$ increases, the temperature increases. A possible way to obtain equilibrium, namely an equal temperature for all the digits, is to use a spectral filter, with nine different frequencies that can be obtained from Eq.(13). An alternative way to obtain an equilibrium state is to keep $\Phi$ constant and to distribute the nine symbols according to a density function $\rho(n)\Phi = \Phi(n)$. Such that,

$$\rho(n_i)\Phi = \ln(1+\frac{1}{n_i}) \qquad (14)$$

The relative distribution of digits in equilibrium is,

$$\varphi(n_i) = \frac{\rho(n_i)\Phi}{\sum_{i=1}^{9}\rho(n_i)\Phi} = \frac{\rho(n_i)}{\sum_{i=1}^{9}\rho(n_i)} \qquad (15)$$

From Eq.(15) it is seen that $\Phi$ disappears altogether, similarly to the way the normalized entropy is independent of the frequency and temperature and became information, in the high occupation limit. Since $\sum_{i=1}^{9}\rho(n_i) = \sum_{i=1}^{9}\ln(1+\frac{1}{n_i}) = \ln(10)$. Therefore,

$$\varphi(n_i) = \log_{10}(1+\frac{1}{n_i}) \qquad (16)$$

Eq.(16) is the Benford's law [16,17,18] that was found empirically in many statistical ensembles of digits that originate from natural sources and are not produced by artificial randomizers.

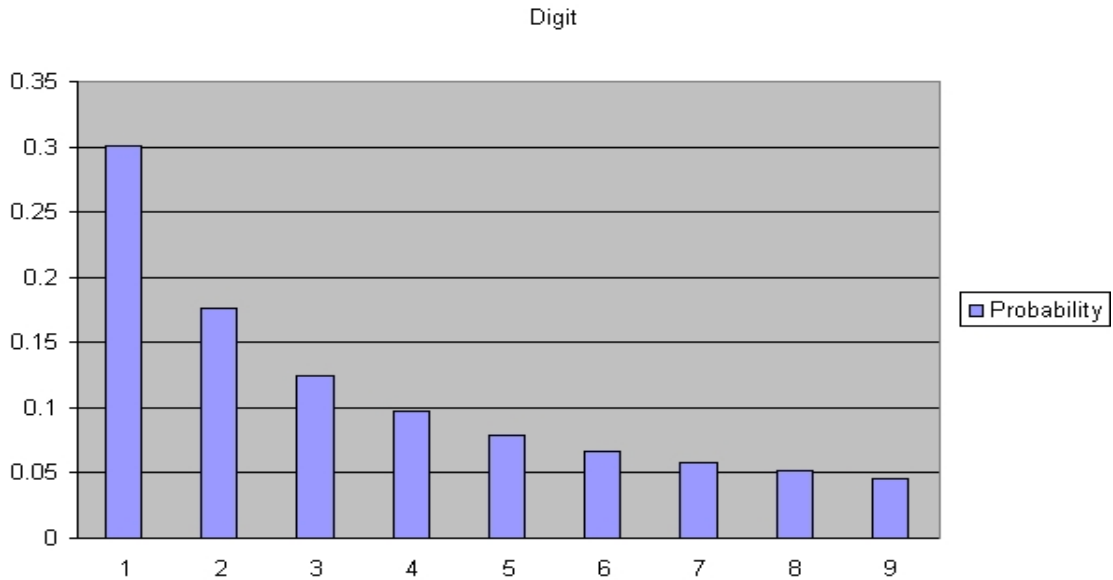

Fig.2 *The Benford's law is the probabilities of the digits in many numerical data files.*

It is worth noting that with the cancellation of Φ we see that the normalized distribution function is independent, not only of the temperature that was chosen arbitrarily in Eq.(13), but also of the energy of the boson as well. Moreover, it is free of any physical variable and/or physical parameter.

**IV-e. Energy distribution – Power-law**

For a canonic ensemble, the energy distribution obtained from the Bose-Einstein statistics is, $n_i = e^{-h\nu_i/k_B T_i}$. The Bose-Einstein distribution gives us the number of photons for any ratio $\Phi = h\nu/k_B T$. Φ may be viewed as the relative energy of a photon with respect to the average energy. Therefore, the occupation number $n$ of all the modes having the same temperature $T$ (in equilibrium) is given by,

$$n_i = \frac{1}{e^{h\nu_i/k_B T} - 1} \qquad (17)$$

The setup that describes Eq.(17) is the same one as in Fig.(1), but without the spectral filer, therefore all frequencies are transmitted. In high occupation number Eq.(17) can be approximate to,

$\Phi_i = \ln(1 + \frac{1}{n_i})$ or $\Phi_i \approx \ln n_i$. We expand $\ln n_i$ around 1 and obtain that $\ln \frac{1}{n_i} \approx \frac{1}{n_i}$ therefore $\frac{\partial \ln \Phi}{\partial \ln n} \approx -1$. In Fig.(2) a plot of $\ln n_i$ Vs. $\ln \Phi_i$ is shown, for in the classical regime a power-law like distribution is obtained, and moreover, the exponential truncation in the quantum regime appears. It is worth noting that the only assumption of this curve is equilibrium. Namely, all modes are at the same temperature [19].

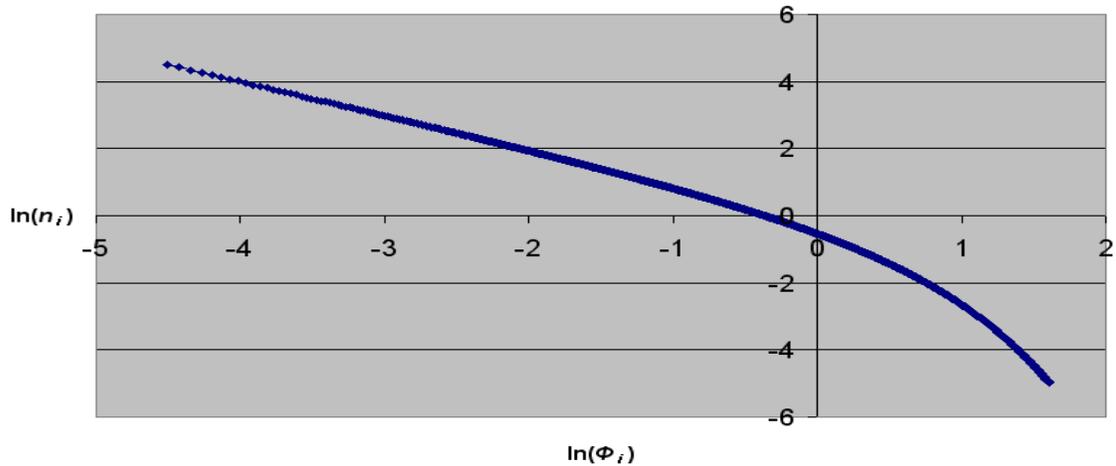

Fig.3 *A log-log plot of the occupation number versus the relative boson energy*

Since many phenomena that are related to natural processes exhibit the power-law distribution [20] it attracts a considerable attention. To mention few: the frequency use of words, the number of hits of web sites, the copies of books sold, the population of cities etc. The slopes obtained from the empirical power-law distributions are around –2. This slope is obtained if we plot the log of the electrical field $E$ instead of the log of energy in the Bose-Einstein distribution (as $\Phi \propto \nu \propto \varepsilon \propto E^2$). This is similar to the Gaussian distribution of momentum that is obtained from the exponential distribution of energy in the canonic limit. A connection of the Bose–Einstein statistics to complex networks was discussed previously [21] and the similarity between the mapping of the Bose-Einstein gas and a network model was pointed out. A possible explanation, based on the present theory, for the reason why so many phenomena exhibit a power-law distribution, is found in section VI.

**V. Combined Systems**

In previous publications [3,4] it was shown that for a classical ensemble the Shannon information is entropy, an amplifier is a Carnot cycle and broadcasting from one antenna to several antennas is a heat flow from a hot bath to a cold bath. In addition, an informatics perpetuum mobile of the second kind was defined. In this paper it is shown that a classical ensemble has a power-law energy distribution while in the quantum limit when $\Phi \gg 1$, the canonic distribution dominates and the regular Gibbs-Boltzmann thermal thermodynamics takes place. Therefore, the Shannon information and the thermal entropy are two faces of the same entropy.

Under what condition thermal entropy will be generated and under what condition information will be generated? When there are two possible ways to generate entropy in a system, the sum entropy will be the combination of the two that maximizes the total entropy of the system [22]. Hereafter a simple example of such a combined system is considered.

**V-a. Hook-law Harmonic Oscillator**

Consider a Hook–law oscillator, at a room temperature $T$, having a spring constant $\kappa$ and an amplitude $A_L$. The total energy of the oscillator is $E_L = \frac{1}{2}\kappa A_L^2$. The temperature of this oscillator from Eq(5) is $T_L = E_L/k_B$. To increase the amplitude of the oscillator to higher amplitude $E_H$, a work $W$ should be applied. The new amplitude will be $E_H \leq E_L + W = \frac{1}{2}\kappa A_H^2 = k_B T_H$. The inequality stands for the situation in which the applied work is not with a resonance with the frequency of the oscillator and therefore part of the work is wasted to heat. It is seen that,

$$\frac{W}{E_H} \leq 1 - \frac{T_L}{T_H}, \qquad (18)$$

Namely, the efficiency of the amplification of the oscillator is the Carnot efficiency. Increasing the energy of the bits in a file was shown [3,4] to be a classical Carnot cycle, which comprises of two isotherms and two adiabatic. Here it is shown that single oscillator amplification is also a Carnot cycle. The Hook oscillator has a weight of a finite mass that affects its frequency. The mass of the weight consists of a large number of particles; each particle has its own degrees of freedom. Each of these particles carries similar entropy to that of the whole Hook-oscillator, which is a single oscillator. Therefore, the temperature of the Hook oscillator is much higher than the thermal temperature of the weight, which is in the room temperature. The Hook oscillator temperature is similar to that of antennas [2] (for a typical cellular antenna was shown to be ~$10^{15}$K) or a laser (for a 0.7µ laser with $10^{16}$ photons per mode is ~$10^{20}$K) and is of the same order of amplitude, namely ~$10^{20}$K. These kinds of temperatures are impossible to obtain by heating up a blackbody by conventional means. Nevertheless, these temperatures can be obtained by non-thermal resonance pumped sources.

Removing energy from the Hook oscillator does not change its entropy because it is a harmonic oscillator and therefore it has a constant entropy $k_B$. However, dumping the oscillator's energy to a canonic ensemble increases the entropy according to Eq(4).

Therefore, the Hook oscillator will dump spontaneously its energy to its thermal bath. This example and similar phenomena are responsible for the common intuition that the information energy is dumped spontaneously into a thermal energy. In fact, this is an example of heat flow from a hot harmonic oscillator to a cold thermal bath.

**V-b. Information versus Thermal Entropy**

Does nature prefers the informatics systems or the thermal canonic systems? This is an interesting question, as we know that our world consists of a mixture of the two. The common intuition, which is based on the canonical thermal physics, suggests a pessimistic end to any closed system, namely, a canonic thermal equilibrium (the heat death that was suggested by Kelvin). The common intuition suggests that informatics is a non-equilibrium phenomenon [23]. Since a file, is a sequence of harmonic oscillators, at the end, the information's energy will relax into a thermal equilibrium exactly as the Hook's oscillator transfers its energy to its bath. However, this is not what the Bose-Einstein statistics suggests. As we see in Fig. 3, there are much more low energy bosons than high-energy bosons.

Eq.(13) suggests that for a given temperature $T$, when $\Phi$ is decreased, $n$ is increased according to,

$$\Phi_i = \ln(1 + \frac{1}{n_i}) \quad (19)$$

When $\Phi <1$, it means that a boson has less energy than the average. When $\Phi >1$ it means that a boson has more energy than the average. Eq(19) suggests that in equilibrium there are more poor energy classical bosons as compared with rich energy (lucky) canonic bosons. A blackbody radiation is a good example of a mixed system. The number of modes in the volume of a blackbody increases with the frequency cube; the wavelength of the light is limited by the diameter of the blackbody. Therefore the occupation number decreases with the frequency according to Eq.(17). The result is the familiar Blackbody radiation spectrum curve that gives similar amount of energy to the poor photons and to the rich photons. Therefore in blackbody emission, the number of the poor photons is much higher than that of the rich photons.

## VI. Summary and Discussion

Based on a toy model, it is shown that the Bose–Einstein distribution, in the quantum limit, yields the regular canonic thermodynamics. In the high occupation limit, the harmonic oscillator statistics replaces the canonic statistics. The harmonic oscillators statistics differs in several aspects from the canonic statistics as is shown in table 1. An important feature of this statistics is that the normalized thermodynamic functions like entropy, and particles distribution do not contain physical quantities. In the canonic entropy the exponential term does not canceled out in the normalization process. Therefore, the canonic entropy is a function of the temperature and the frequency. Any fluctuation of the energy and/or the temperature affects its magnitude. In the high occupation limit entropy, all the physical variables and parameters disappear and we obtain the Shannon information. Therefore, the entropy is not sensitive to any fluctuation in the occupation number, the source temperature and/or frequency. It is not even sensitive to the number of modes in a bit. This property of the entropy, in the high occupation limit, makes it appropriate to convey data.

| | High Occupation $n \gg 1$ | Canonic $n \ll 1$ |
|---|---|---|
| Temperature | $T = \dfrac{nh\nu}{k_B}$ | $T = -\dfrac{h\nu}{k_B \ln n}$ |
| Equilibrium | $p = 1/2$ | $\dfrac{p}{1-p} = e^{-\frac{h\nu}{k_B T}}$ |
| Average mode entropy | $S = k_B \ln 2$ | $S = \dfrac{h\nu}{T} e^{-\frac{h\nu}{k_B T}}$ |
| Energy Distribution | Power-law | Exponential |
| Carnot cycle | Amplifier | Heat engine |

Table 1 *The thermodynamic properties of the Bose-Einstein gas in equilibrium at temperature T for photons (with zero chemical potential) for the classical and the canonic distribution. p is the probability of the energetic modes and n is the occupation number.*

The logical quantities, in the high occupation limit, are therefore applicable to many phenomena of our life. The Bose-Einstein distribution of photons is a simple combinatory of states and particles without interactions, as the chemical potential µ=0. The only constraint encapsulate in it is the quantization. Namely, it is possible to add or to remove energy from any mode in an integer amount of some undivided particle (a quant). As is seen in Eq.(15) when we keep the quant size fixed (a constant frequency) and we also assumed equilibrium (equal temperature for all the modes), than the normalized distribution of the photons is not a function of the energy, the frequency, the average energy or the temperature. The physics is faded away, and we remain with a statistical system of inert quanta. Systems like these are very common in life. Consider the distribution of the population of cities. Each city may be considered as a mode. When we count the number of the peoples in a city, the peoples are, per definition, indistinguishable. Since the number of the peoples is quantized, therefore this system is identical to that of Eq.(15). Similarly, the number of books being sold in a certain period of time is a homological system to that of the population of cities. In this case the number of the titles is the number of the modes and a single copy sold is a quant. The number of hits in the Internet is also a system of this kind as the number of the sites is the number of modes and a hit is a quant.

In the derivation of Benford's law Eq.(14) was used, namely, $\rho(n_i)\Phi = \ln(1+\frac{1}{n_i})$. This equation yields slope of "-1". In the normalization process $\Phi$ disappears. A slope "-2" is obtained if we substitute $\psi^2(n) = \rho(n)$, with a phenomenological analogy to the substitution of momentum instead of energy in canonic exponential distributions to obtain the Gaussian distribution.

The present model does not consider any interactions between the quantized particles. Nevertheless, interactions do exist. If we consider, for example, the distribution of the hits among the sites in the Internet, it is obvious that there are interactions between the visitors of the sites. The interactions might be advertisements by the sites and/or viral spread of the recommendations by the visitors. So what is the reason for a somewhat oversimplified model without interactions being so effective? A possible explanation is that the distribution of the hits is independent of the interactions, however a specific rank of a certain site does depend on the interactions. Namely, the interactions are responsible only for the specific site location in the distribution. If that is true, removing a several popular sites will not change the normalized distribution. Other sites will take the place of the removed sites and the distribution will reach equilibrium again. Indeed, this is what is seen in almost any economical system, namely "there is no empty space". Unlike the derivation of Benford's law, the present model does not pretend to be a complete solution to the power-law distribution in social systems. Nevertheless, it is argued that extensive equilibrium thermodynamics may predict the qualitative behavior of social systems.

Another notable property of the logical equilibrium is the quenched randomness. For the receiver, a random file is content. However, within the context of IT, a random file, which is a compressed file, is an ensemble of harmonic oscillators in equilibrium, as is seen in Eq(12). An outcome of this conclusion is that files should have a tendency to be compressed spontaneously. Aside from the natural spontaneous noise, we are obsessed with compression. In IT we compress files for economical reasons. However, an observer in space sees that most of the transmitted files on earth are compressed. This observer will rightly, conclude that files have a tendency to be compressed. Our tendency to compress is seen also in art. We find ourselves impressed by an artist who can express a complex feeling with a few words, or by a painter who can represent a detailed picture with a few lines and colors. The artistic kind of compression is known in IT as a lossy compression and is very popular in multimedia technology. The language is a most powerful compressor; sometimes, the amount of information in a short sentence is enormous considering the fact that it contains just a few bytes. A notable example is the mathematics, which enables to write relatively short formulas that describe complex logical processes. It is possible that our tendency for symbolism and mathematics is the natural tendency toward equilibrium.

The last issue and the most intriguing one is how the tendency of information to increase affects life. Conventional canonic thermodynamics explains how we decompose chemical compounds in order to produce mechanical work, and heat to enable our body to function properly. This paper suggests that we also want to increase information. The increase of information can be done by reproduction and by broadcasting. It is clear that the present evolution theories are with full agreement with the present theory [22]. The only modification required is that reproduction and evolution are spontaneous processes. It was shown previously [2,3] that information is multiplied in broadcasting. Therefore, it is not surprising that we are obsessed with a desire to broadcast ourselves. When Bob broadcasts a file with $I$ bits to $N$ receivers, he will increase the information by $NI$. A receiver will increase the information by $I$ bits. Therefore, it is better, thermodynamically, to broadcast than to receive.

It is an observable fact that information and life in their various forms increase with time; therefore, it is plausible that aside from the chemistry necessary for the existence, life means a reproduction and a compressed communication.

**Acknowledgments**: I thank J. Agassi, Y. B. Band, Y. Kafri, H. Kafri, the late O. Meir and R. D. Levine for many discussions throughout this work.